\documentclass[aps,preprint,nofootinbib]{revtex4}
\usepackage{graphicx}
\usepackage{color}
\usepackage{epsfig}

\begin{document}
\title{Liquid-liquid critical point in supercooled silicon}

\author{Vishwas V Vasisht}
\author{Shibu Saw}
\author{Srikanth Sastry\footnote{Corresponding Author: sastry@jncasr.ac.in}}

\affiliation{Jawaharlal Nehru Centre for Advanced Scientific Research, Bangalore
560064, India.}

\maketitle 

{\bf A novel liquid-liquid phase transition has been proposed and
  investigated in a wide variety of pure substances recently,
  including water, silica (major components of the earth's crust), the
  technologically important element silicon, and others marked by
  energetically stabilized tetrahedral local geometries. From computer
  simulations using a classical empirical potential (the
  Stillinger-Weber potential), Sastry and Angell
  \cite{sastry_angell_si} demonstrated a first order liquid-liquid
  transition in supercooled silicon, supported by further experimental
  and simulation studies subsequently. Here, we report evidence for a
  critical point to the liquid-liquid phase transition at negative
  pressures, from computer simulations using the SW
  potential. Compressibilities, evaluated from the equation of state
  and fluctuations in constant pressure-temperature simulations
  exhibit a growing maximum upon lowering temperature below $1500 K$
  and isotherms exhibit density discontinuities below $1120 K$, at
  negative pressure. Below $1120 K$, isotherms obtained from constant
  volume-temperature simulations exhibit non-monotonic, van der
  Waals-like behavior signaling a first order transition.  We identify
  $T_c \sim 1120 \pm 12 K$, $P_c \sim -0.60 \pm 0.15 GPa$ as the
  critical temperature and pressure for the liquid-liquid critical
  point.  The structure of the liquid changes dramatically upon 
  decreasing the temperature and pressure. Diffusivities vary over 4 orders of magnitude, and
  exhibit anomalous pressure dependence near the critical point. A
  strong relationship between local geometry quantified by the
  coordination number, and diffusivity, is seen, suggesting that atomic
  mobility in both low and high density liquids can usefully be
  analyzed in terms of defects in the tetrahedral network
  structure. We have constructed the phase diagram of supercooled
  silicon.  We identify the lines of compressibility, density extrema
  (maxima and minima) and the spinodal which reveal the
  interconnection between thermodynamic anomalies and the phase behaviour of the
  system as suggested in previous works \cite{speedy_angell,speedy_conjecture,debenedetti_spinodal_anomalies1,poole_water,sastry_singularityfree,stanley_mishima,debene2009,saika_voivod_silica}. }


The possibility of a phase transition between two forms of the liquid phase 
in some pure substances has attracted considerable interest and research activity 
in recent years \cite{sastry_angell_si,speedy_angell,speedy_conjecture,poole_water,stanley_mishima,debene2009,saika_voivod_silica}. Among the substances investigated are water \cite{poole_water,stanley_mishima,debene2009},
silica \cite{saika_voivod_silica} and silicon \cite{sastry_angell_si,aptekar,donovan,deb_sli_exp,helder_lsi_exp,borick,ashwin,Ganesh_FPMD_si}, germanium, carbon and hydrogen -- these substances together 
form a very significant component of our natural world, living organisms, and technology. 
A phenomenon common to these is therefore of wide general interest. As illustrated in 
\cite{kurita_fluidity}, further, liquid-liquid transitions offer an avenue for 
interesting applications by exploiting the different properties of distinct liquid phases.

Although the liquid-liquid transition was discussed in the context of
silicon \cite{aptekar} somewhat earlier, the considerable current
interest stems from various proposals for understanding the anomalies
of water
\cite{speedy_conjecture,poole_water,sastry_singularityfree,stanley_mishima,debene,chen_confined_water,caa_critfree}. These
scenarios have alternately invoked the approach to a
spinodal\cite{speedy_conjecture}, a liquid-liquid critical
point\cite{poole_water,stanley_mishima}, general thermodynamic
constraints without the presence of any singular
behavior\cite{sastry_singularityfree}, and the presence of a
transition without a critical point\cite{caa_critfree}, in
rationalizing experimentally observed behavior.  In spite of
substantial investigations, a general consensus is yet lacking on the
interpretation of observed behavior \cite{debene,caa_critfree}. In particular,
recent experiments on confined water \cite{chen_confined_water} and
issues surrounding their interpretation \cite{caa_critfree} indicate
the need to ascertain the existence of a critical point even when
sufficient evidence exists for a liquid-liquid transition.

The possibility of a transition in supercooled silicon was suggested
\cite{donovan} based on estimates of excess Gibbs free energies of
amorphous and liquid silicon, implying a ``amorphous-liquid'' phase
transition near $1,450$ K (below the freezing point of the liquid,
1,685 K).  Clear evidence of a transition between two liquids in the
supercooled region was shown in \cite{sastry_angell_si} from molecular
dynamics simulations using Stillinger-Weber (SW) potential
\cite{still_weber_pot}. A first order transition at zero pressure was
found at $T$ = $1060$ K, substantially below the experimental
estimate. However, recent electrostatic levitation experiments
performed down to $T = 1382 K$ do not find evidence for a transition
\cite{kelton}. Apart from the uncertainties in the experimental value,
such a difference may be expected to arise from the neglect in the
empirical potential of significant changes in the electronic structure
associated with structural change \cite{ashwin, Ganesh_FPMD_si},
though first principles simulations \cite{Ganesh_FPMD_si} appear to
confirm the transition temperature obtained from the classical
simulations.  The exact location of the transition must thus be viewed
as tentative at present.

\begin{figure}[h]
\begin{center}
\epsfig{file=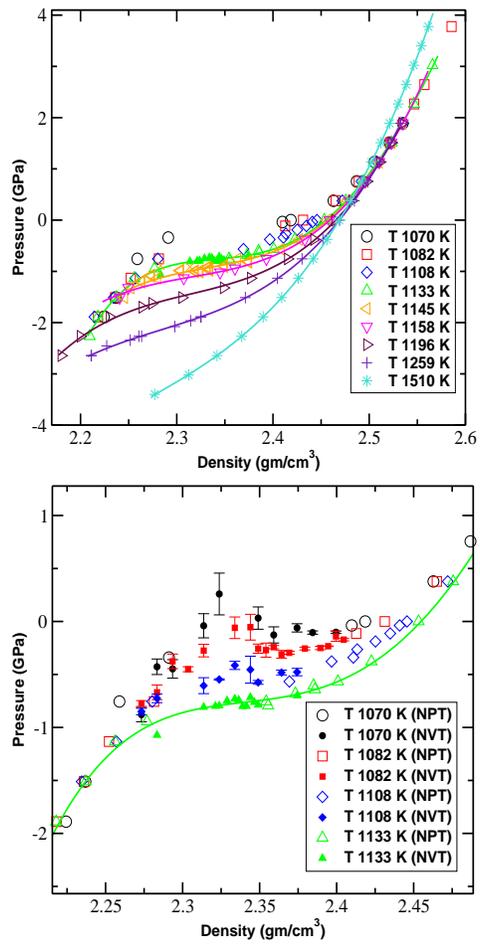,scale=0.7,angle=0,clip=}
\caption{\bf Vasisht et al}
\label{fig:EOS}
\end{center}
\end{figure}

In the present work, we use molecular dynamics (MD) simulations using
the SW potential for silicon to locate the liquid-liquid critical
point, and show that it lies at negative pressures. We perform
constant pressure, temperature (NPT) and constant volume, temperature
(NVT) simulations of $512$ atoms, using protocols described in {\it
  Methods}.  Pressure {\it vs.} density isotherms generated in the
temperature range of $1070K$ to $1510K$, and the pressure range of
$-3.8GPa$ to $+3.8GPa$ using NPT simulations are shown in
Fig. \ref{fig:EOS} (top panel). The isotherms for temperatures above
$T = 1133 K$ are continuous, but develop an inflection below $T = 1259
K$ which becomes more pronounced as temperature is lowered. The
compressibility develops a maximum in this temperature range, which
grows as the temperature is lowered.  Below $T = 1108 K$, careful
constant pressure simulations always result in a jump in the density
as pressure is varied, suggesting a first order transition.  To verify
this further, we perform NVT simulations for $T = 1108 K, 1082 K, 1070
K$ in the density range where NPT simulations show a jump. These
results are shown in the bottom panel of Fig. \ref{fig:EOS}. We find
that at these temperatures, the isotherms are non-monotonic. Such
non-monotonicity in simulations arises from metastability on the one
hand, and on the other hand, incomplete phase separation owing to
finite sample sizes in the unstable region, and constitutes a clear
indication of a first order transition. Thus, our equation of state
data show isotherms with growing compressibility maxima as temperature
is decreased (above $T = 1133 K$) and first order transitions (below $T
= 1108 K$) between two liquids, the high density liquid (HDL) and the
low density liquid (LDL). We thus deduce that the critical point
to be located between these temperatures. Inspection of
Fig. \ref{fig:EOS} (bottom panel) also makes it clear that the
critical pressure must be negative. Based on the above data at the
temperatures simulated, we estimate the location of the critical point
to be at $T_c \sim 1120 \pm 12 K$, $P_c \sim -0.6 \pm .15 GPa$. A more
precise estimation of critical parameters require analysis, including
finite size scaling, which is beyond the scope of the present paper
but are being pursued as extensions of the present work.

\begin{figure}[h]
\begin{center}
\epsfig{file=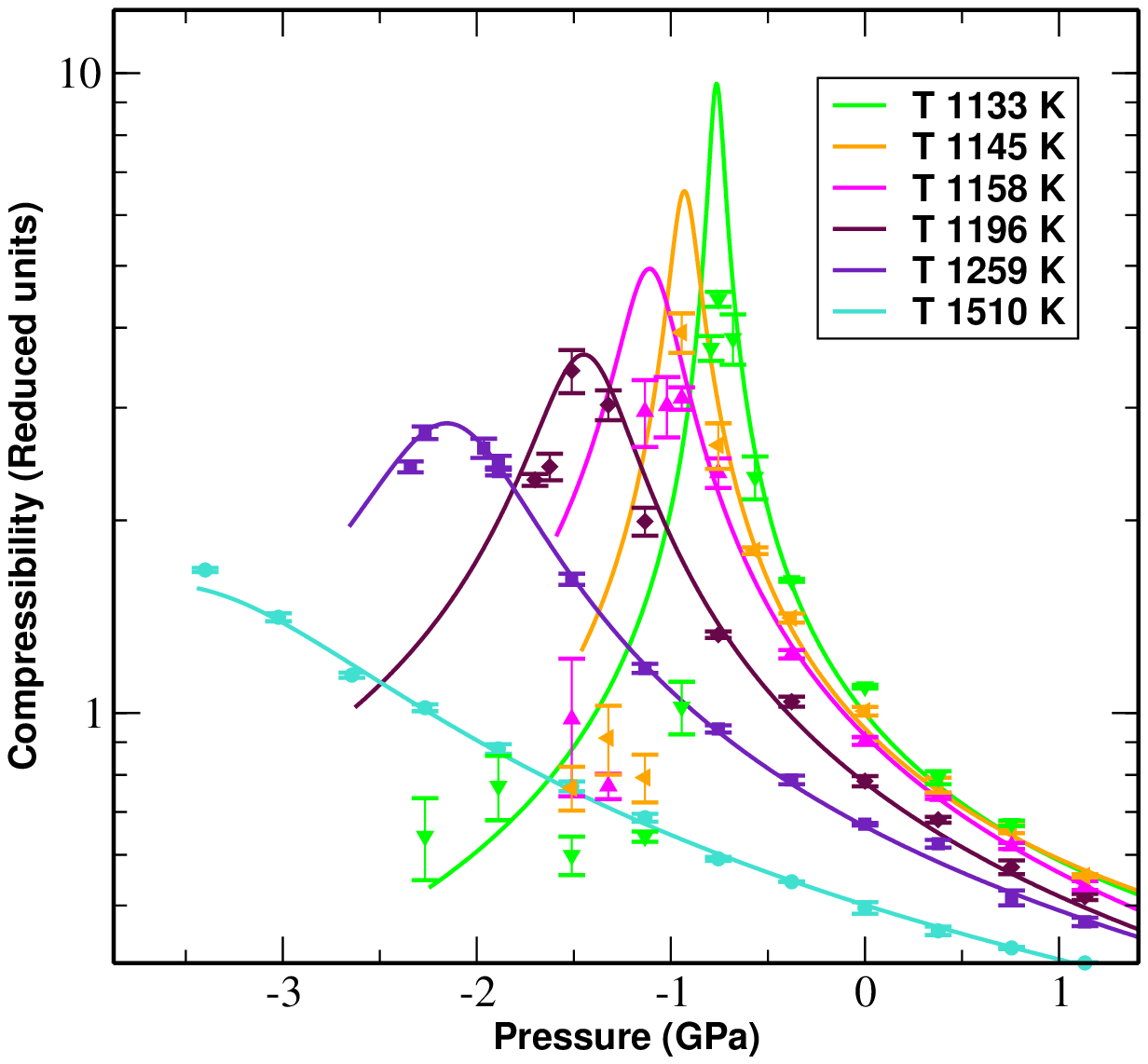,scale=0.7,angle=0,clip=}
\caption{\bf Vasisht et al}
\label{fig:Compressibility}
\end{center}
\end{figure}

Approaching the critical point from above leads to increased density fluctuations. In addition to 
evaluating the compressibility from the equation of state (EOS), we also calculate it directly from 
density fluctuations. These are shown in Fig. \ref{fig:Compressibility} for temperatures above 
$T = 1133 K$ which show good agreement with EOS estimates for the high density liquid (HDL), but 
poorer agreement for the low density liquid (LDL). The high crystallization rates observed near and 
at lower pressures than the compressibility maxima hamper improved sampling (however we report 
results only from equilibrated runs). The influence of fluctuations and local structure (see below) 
on nucleation rates \cite{daan1,daan2} is an intersting issue that is presently being investigated. 

\begin{figure}[h]
\begin{center}
\epsfig{file=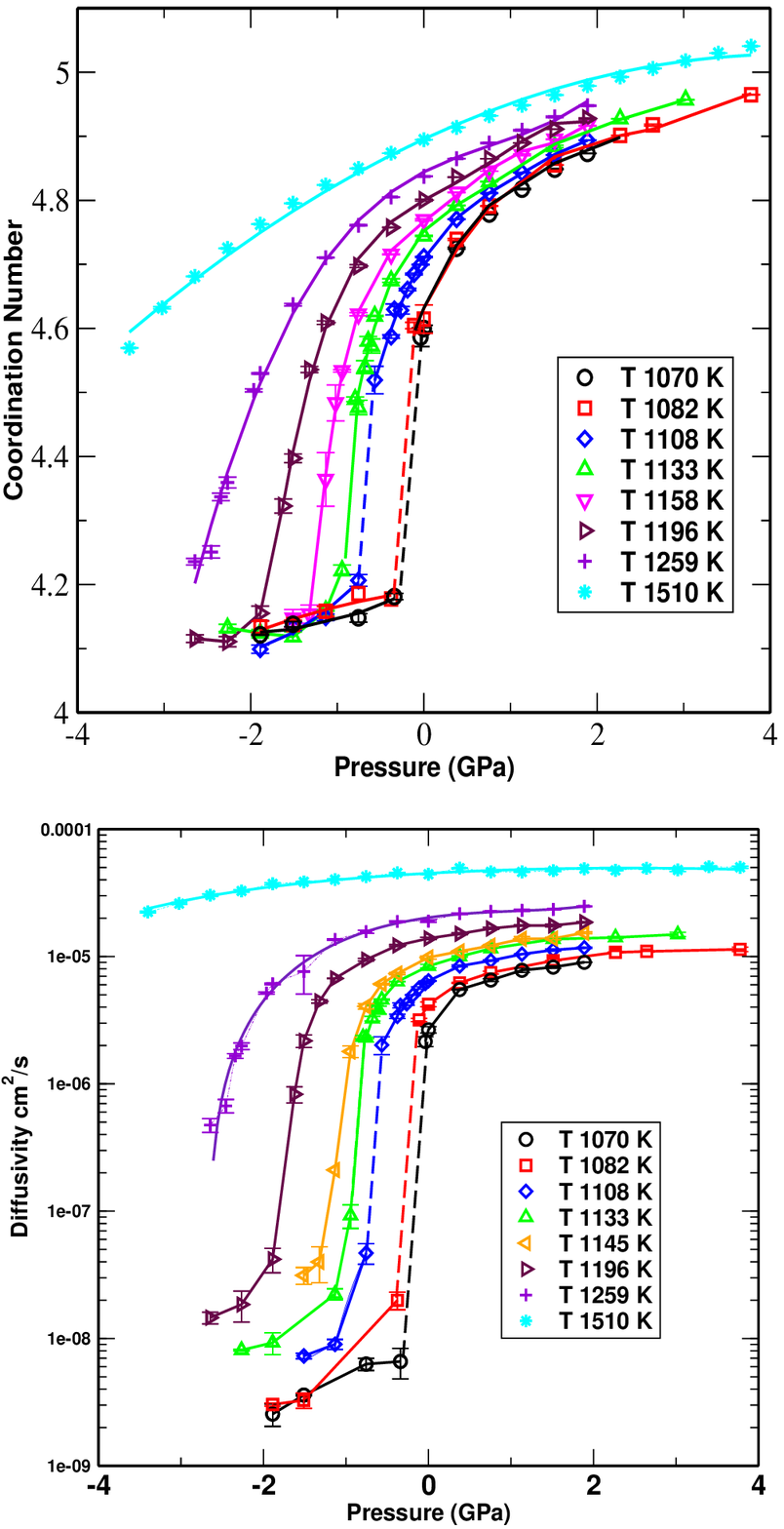,scale=0.7,angle=0,clip=}
\caption{\bf Vasisht et al}
\label{fig:Coord_Diff_Pres}
\end{center}
\end{figure}

Next we describe briefly the temperature and density dependence of
diffusivities $D$ and average coordination number $N_n$, which captures
important information regarding structural change. $N_n$, 
 the {\it number of neighbors} of an atom is 
calculated by integrating the pair correlation function up to its
first minimum.  In Fig. \ref{fig:Coord_Diff_Pres} (top panel) $N_n$ 
is shown as a function of pressure. At high temperatures and
pressures, $N_n$ is about $5$, and decreases as
temperature and pressure are lowered, showing discontinuous change
below $1133 K$, and values close to the tetrahedral value of $4$
($\sim 4.1$) in the LDL, similar to the observation at zero pressure
in \cite{sastry_angell_si}.

Diffusivities {\it vs.} pressure for the studied temperatures are
shown in (Fig. \ref{fig:Coord_Diff_Pres} - bottom panel), which
increase with pressure for all the temperatures shown, in analogy with the well known anomaly
in water. Like the coordination number, diffusivities show a
discontinuity below $T = 1133 K$, with a jump of roughly two orders of
magnitude from HDL to LDL. The diffusivities span a range of over four
orders of magnitude.

\begin{figure}[h]
\begin{center}
\epsfig{file=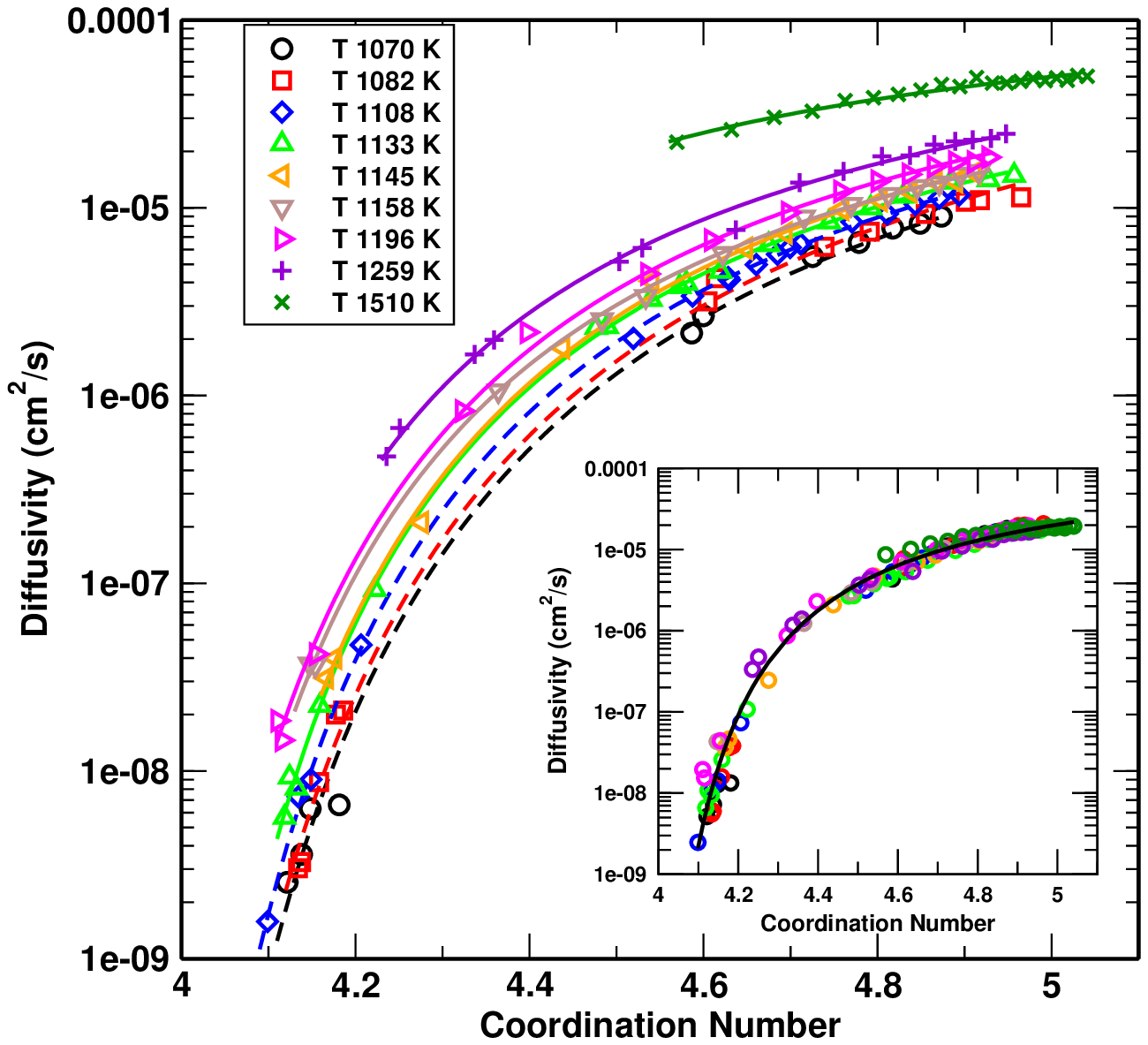,scale=0.7,angle=0,clip=}
\caption{\bf Vasisht et al}
\label{fig:Diff_Coord}
\end{center}
\end{figure}

In Fig. \ref{fig:Diff_Coord}, we show the diffusivities $D$ plotted
against coordination number $N_n$. Except at the highest temperature
studied, we find the dependence of $D$ on $N_n$ to be remarkably
similar irrespective of temperature, including those below the
critical temperature. The mobility of atoms therefore is very strongly
determined by the local structure. This observation is consistent with
previous analysis of the role of ``bifurcated bonds'' or the ``fifth
neighbor'' in determining molecular mobility in water
\cite{sciortino_mobility}. It is tempting to speculate that apart from
trivial thermal effects, a universal dependence exists of atomic
mobility on the average number of neighbors in excess of the
tetrahedral value of $4$. To test this possibility, we show in the
inset of Fig. \ref{fig:Diff_Coord} a scaled plot of $D$, by
normalizing to its value at a fixed $N_n$ in the HDL phase for all
temperatures. The data collapse is indeed remarkable and the resulting
master curve can be well fit to a Vogel-Fulcher-Tammann (VFT) form,
$D(n) = D_0 exp(-\frac{A}{n - n_0})$ with $n_0 = 3.86$. Further
analysis of this interesting observation is in progress.

\begin{figure}[h]
\begin{center}
\epsfig{file=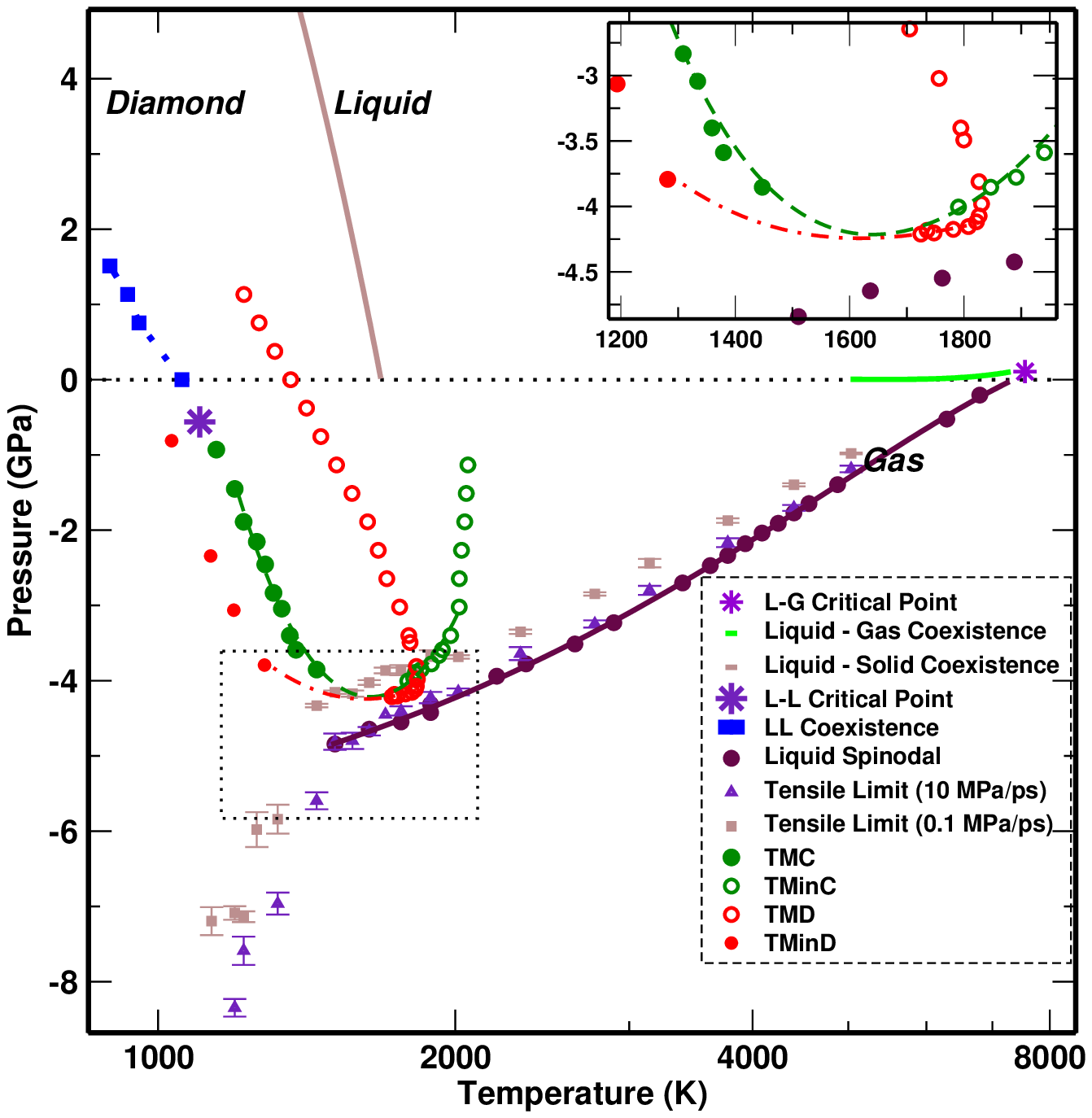,scale=0.7,angle=0,clip=}
\caption{\bf Vasisht et al}
\label{fig:Phase_Diagram}
\end{center}
\end{figure}

Previous studies of the phase behavior of water and other liquids
exhibiting density anomalies have analyzed the interplay of various
loci of extremal behavior, namely the spinodal, lines of density
maxima (TMD) \cite{speedy_conjecture}, density minima (TMinD)
\cite{denmin}, compressibility minima (TMinC) and maxima (TMC).  To
obtain a comprehensive picture of the phase behavior of liquid
silicon, we have evaluated these loci, employing in addition to the MD
simulations, parallel tempering (PT) and restricted ensemble (REMC)
Monte Carlo simulations \cite{corti_remc} for locating the spinodal
(details and data are provided in the supplementary information (SI))
at low temperatures.  As shown in Figure \ref{fig:Phase_Diagram}, the
spinodal we estimate is monotonic in pressure {\it vs.} temperature
$T$, {\it ie} not ``reentrant'' as predicted to be the case
\cite{speedy_conjecture} if it intersects with the TMD. The TMD,
however, changes slope upon intersection with the TMinC, as analyzed
in \cite{sastry_singularityfree}. Evaluating the relevant equation of
state data is particularly challenging in this case as the TMD approaches the 
spinodal very closely, while not intersecting it. From available data, it appears that
the TMinC will join smoothly with the TMC (line of compressibility
maxima) that emanates from the liquid-liquid critical
point. Interestingly, we find from PT simulations below the critical
temperature and pressure that there exists also a line of density
minima, very recently observed in the case of water in experiments and
computer simulations \cite{denmin}. The TMD and the TMinD appear to
smoothly join with each other, as required by thermodynamic consistency. 

At low temperatures $T$ ($1440 K < T < 2000 K$), even REMC simulations
(which restrict density fluctuations) cavitate at sufficiently low
pressure. In these cases, we estimate the spinodal by a quadratic
extrapolation of the isotherms. As a further check on our spinodal
estimate, we perform simulations to obtain tensile limits by
increasing the tensile pressure on the simulation cell at constant
rates, for two different rates (0.1 and 10.0 MPa/ps).  For the faster
rate, we find tensile limits that are consistent with the spinodal
estimates we have, while for the slower rate, the system cavitates at
higher pressures, remaining nevertheless monotonic {\it vs.}
temperature. These simulations also extend our estimate of the
spinodal to lower temperatures, and indicate a marked downturn of the
spinodal pressure below the temperatures we have studied.

In conclusion, we have performed extensive molecular dynamics and
Monte Carlo simulations of supercooled liquid silicon using the SW
potential to find evidence for a negative pressure liquid-liquid
critical point.  We estimate the location of the critical point to be
at $T_c \sim 1120 \pm 12 K$, $P_c \sim -0.60 \pm 0.15 GPa$.  We have
computed the phase diagram of supercooled liquid silicon. We find no
retracing of spinodal. The phase behaviour of silicon is similar to
that observed in simulations of water and silica. The structure of the
liquid changes dramatically in going from high temperatures and
pressures to low temperatures and pressures. Diffusivities vary by
more than 4 orders of magnitude, and exhibit anomalous pressure
dependence.  A strong relationship between local geometry quantified
by the coordination number and diffusivity is seen, suggesting that
atomic mobility in both low and high density liquids can usefully be
analyzed in terms of defects in the tetrahedral network structure.



\noindent{\bf Methods :}

We perform Molecular Dynamics (MD), with a time step of 0.383 fs, with
details as in \cite{sastry_angell_si}, but employing an efficient
algorithm\cite{shibu_sw} for energy and force evaluations.  NVT
simulations employ the LAAMPS \cite{lammps} parallelised MD package.
PT and REMC Monte Carlo simulations are described in the SI.

In the HDL phase, a mininum of 3 to 6 independent samples are
simulated for $\sim 100$ relaxation times ($\sim$ $10$ $ns$). In the
LDL phase, crystallisation (monitored by energy jumps, mean square
displacement (MSD) and pair correlation function) rates are high. We
perform around 10 to 50 initial runs, each of $22$
ns. Noncrystallizing samples (average of 5) were run up to 10
relaxation times when possible. In all LDL cases, simulations were
carried out for times required for the MSD to reach $1 nm^2$ (5
$\sigma^2$, where $\sigma$ is the atomic diameter) or for $100 ns$
(300 million MD steps), whichever is larger.

Equilibration is monitered by the MSD and the overlap function Q(t),
defined as $Q(t) = \sum_{i=1}^{N} w|\vec r_i(t_0) - \vec r_i(t +
t_0)|$, where $w(r)$ = $1$, if $r \leq 0.3\sigma$, zero otherwise.  We
evaluate relaxation times ($\tau$) by fitting $Q(t)$ to a stretched
exponential function. As an indication of its variation, $\tau$ varies
at $P = 0 GPa$ in the HDL from $0.3$ ps (T = $1260$ K) to $0.01$ ns (T
= $1068$ K) and in LDL phase at T = $1060$ K, the $\tau$ is around
$30$ns.

We calculate the compressibility $K_T$ from the EOS using: 

$$
     K_{T} = \frac{1}{\rho} \left [ \frac{\partial \rho}{\partial P} \right ]_{T}.
$$

Polynomial fits to isotherms are used in calculating the
derivatives. We also calculate $K_T$ from volume fluctuations (NPT
simulations) using:

$$
     K_{T} = \frac{\left < V^{2} \right > - \left < V \right >^{2}}{\left < V \right >k_B T}.
$$

The latter method is computationally very demanding, and the
comparison between the two reveals degree to which sampling is
satisfactory. In HDL the two estimates of $K_T$ agree very well, but
in LDL, below the $K_T$ maximum, the deviations between the two
indicate that sampling in LDL is not sufficient to obtain $K_T$ from
fluctuations. 


\eject 




\centerline{\LARGE \bf Figure Captions}

{\bf Figure 1: Equation of state.} (top panel) Plot of pressure
against density isotherms from NPT simulations.  Isotherms for $T \ge
1133 K$ are continuous, and the lines through the data points are
polynomial fits used to obtain the compressibility. For $1133 K \le T
\le 1259 K$, the isotherms show an inflection, corresponding to a
compressibility maximum. For $T < 1133 K$, the isotherms show a
density discontinuity signaling a first order transition. The crossing
of isotherms at positive pressures reflects the presence of density
maxima. (bottom panel) Pressure {\it vs.} density below the critical
point from NPT (open symbols) and NVT (filled symbols)
simulations. Pressures from NVT simulations below $T = 1133 K$ exhibit
non-monotonic behavior at intermediate densities indicating a phase
transition, while matching with pressures from NPT simulation at high,
low densities. The critical temperature lies between $T = 1133 K$ and
$T = 1108 K$.

{\bf Figure 2: Compressibility maxima increase with decrease in
  temperature.} Compressibilities $K_T$ {\it vs.} pressure for
different temperatures from NPT simulations. The lines show $K_T$
calculated from the derivative of the pressure along isotherms and the
symbols show $K_T$ calculated from volume fluctuations. The maximum
value of $K_T$ increases as temperature decreases, indicating an
approach to a critical point.

{\bf Figure 3: Coordination number and Diffusivity.} (top panel) Plot
of coordination number $N_n$ against pressure for different
temperatures obtained from NPT simulations. In HDL, the $N_n$ is
around five and in LDL it approaches four, corresponding to
tetrahedral local structure. (bottom panel) Plot of diffusivity $D$
against pressure for different temperatures. $D$ is calculated
from the mean square displacement, and shows a dramatic drop
of over two orders of magnitude as the liquid transforms from HDL to LDL.

{\bf Figure 4: Relationship between structure and dynamics.} Plot of
diffusivity $D$ against coordination number $N_n$ at different
temperatures.  Lines through the data points are guides to the eye,
and highlight the remarkably similar dependence of $D$ on $N_n$ at all
temperatures including those below the critical temperature where both
$D$ and $N_n$ change discontinuously. (inset) Plot of $D$ (scaled to
match at $N_n = 4.8$) {\it vs.} $N_n$, showing data collapse. The line
is a Vogel-Fulcher-Tammann (VFT) fit, with a coordination number of
vanishing diffusivity $ = 3.86$.

{\bf Figure 5: Phase Diagram in PT plane.} The pressure-temperature 
phase diagram showing the location of 
(a) the liquid-crystal phase boundary \cite{voronin},
(b) the liquid-gas phase boundary and critical point,
(c) the liquid-liquid phase boundary and critical point,
(d) the liquid spinodal 
(e) the tensile limit obtained from two different increasing tensile pressure at two different rates,
(f) the density maximum (TMD) and minimum (TMinD) lines, and 
(g) the compressibility maximum (TMC) and minimum (TMinC) line.  
Lines joining TMD and TMinD (dot-dashed), TMC and TMinC (dashed) are guides to the eye. 
(inset) Enlargement of box in main panel showing that TMD and spinodal lines do not intersect.

\eject

\centerline{\LARGE \bf Figures}

\pagebreak

\pagebreak

\pagebreak 

\pagebreak 

\end{document}


\begin{center}
\textbf{Supplementary Information}
\\
\textbf{Liquid-liquid critical point in supercooled silicon}
\\
\textbf{Vishwas V Vasisht, Shibu Saw, Srikanth Sastry}
\end{center}

Here we present additional details of simulations that are performed
to construct the phase diagram of liquid silicon. These include the
estimation of the liquid-liquid coexistence line, the spinodal, the
tensile limit loci, the lines of density maxima and minima, and the
lines of compressibility maxima and minima. 

\begin{enumerate}
	\item {\bf Line of Liquid-Liquid Coexistence: }
	      The coexistence line is calculated from isobars generated using NPT MD simulations (with system size of 512 particles). 
	      We find sudden jumps in density for small changes in temperature, which indicate the location of the first order phase transition.

	      \begin{figure}[h]
		      \begin{center}
			      \epsfig{file=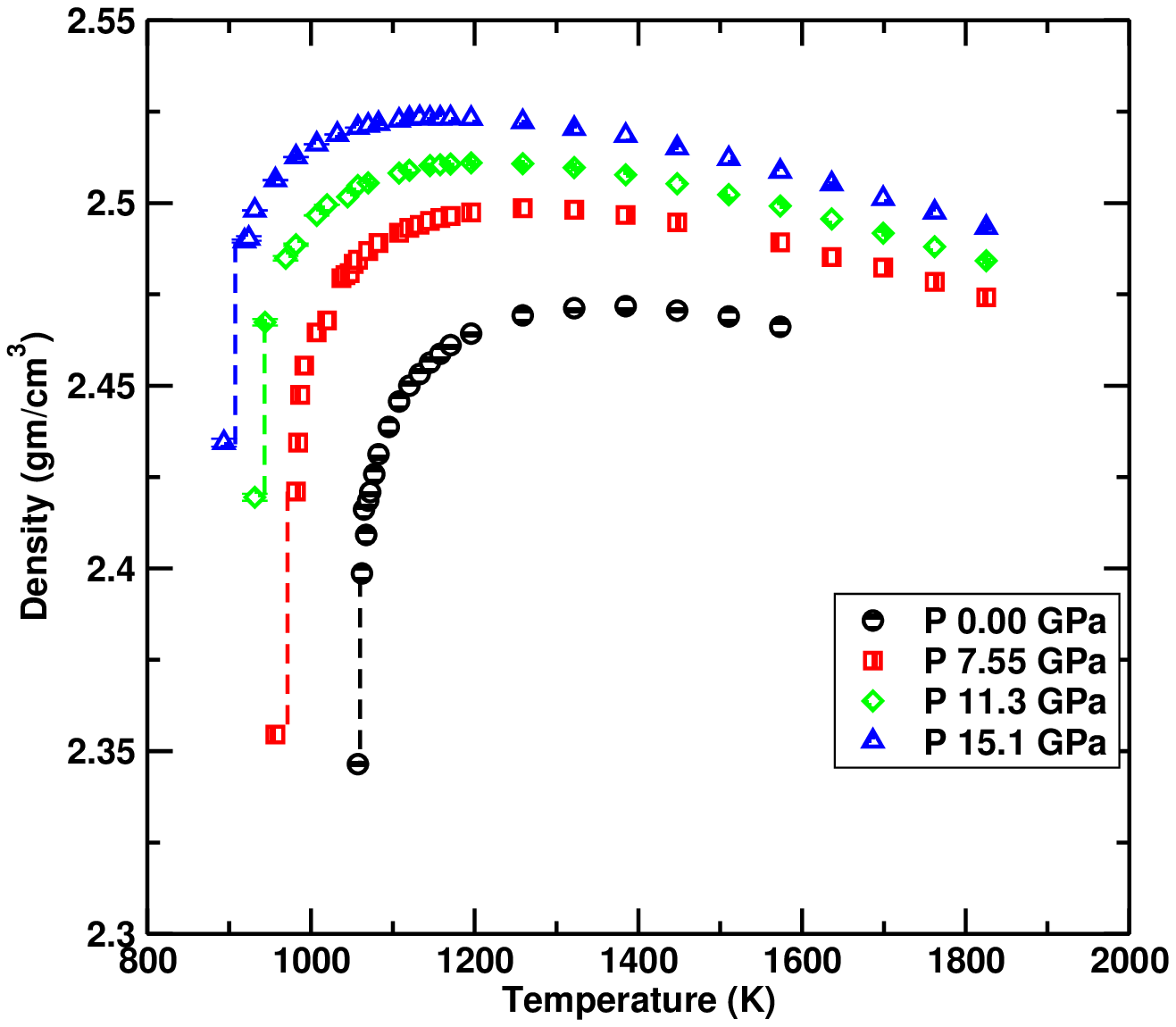,scale=0.7,angle=0,clip=}
			      \caption{Density {\it vs.}
                                temperature isobars showing
                                discontinuous changes in density for
                                small changes in temperature at
                                constant pressure. These jumps are
                                used to locate the temperature at
                                which a first order transition occurs
                                at the studied pressures.}
			      \label{fig:Coexistence}
		      \end{center}
	      \end{figure}

      \item {\bf Liquid-Gas coexistence line: } To obtain the L-G coexitence line we perform Gibbs-Ensemble Monte Carlo - GEMC 
	({\it see Understanding Molecular Simulation by D. Frenkel and B. Smit}) simulations (with 2000 particles). In a 
	      GEMC simulation we obtain, at coexistence, liquid and gas phases in two different simulation boxes. When phase equilibrium
	      is achieved, the pressure, the temperature and the chemical potential of both the simulation boxes will be the same. At each 
	      temperature we obtain the coexistence pressure to construct the liquid-gas coexistence line.

      \item {\bf Spinodal line: } The spinodal point is defined by the condition
	      $(\partial P/\partial V)_T = 0$.  In Fig. \ref{fig:HighT_Spinodal} we show 
	      high temperature spinodal isotherms ($T > 2200 K$). These isotherms are obtained
	      from constant volume temperature (NVT) MD simulations, with system size of 512 particles. 
	      At each state point we perform simulations for around $7$ ns to $20$ ns. 
	
	      \begin{figure}[h]
		      \begin{center}
			      \epsfig{file=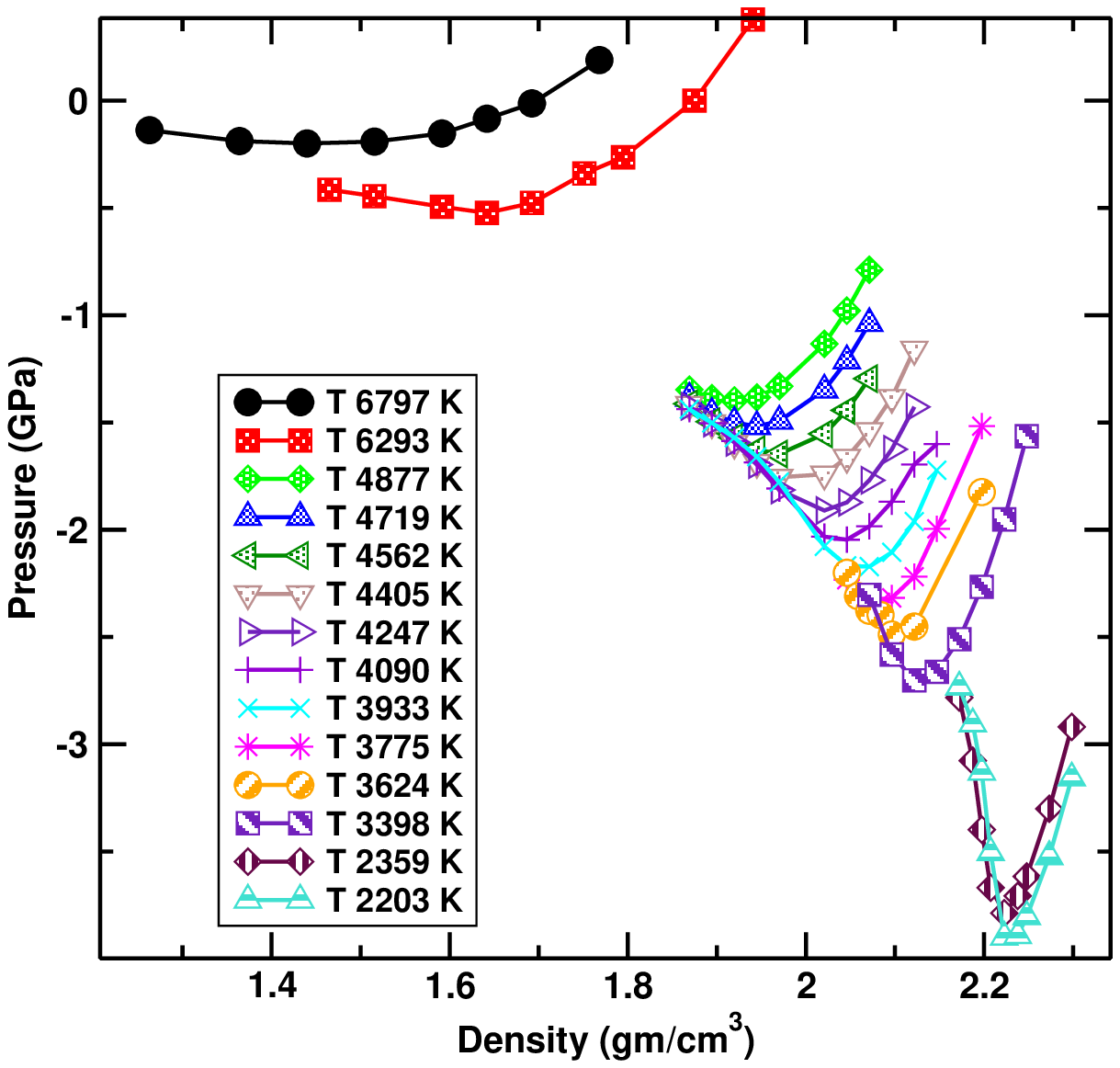,scale=0.5,angle=0,clip=}
			      \caption{The plot of pressure {\it vs. } density  showing high 
			      temperature isotherms. The isotherms go through minima which are identified as 
			      spinodal points.}
			      \label{fig:HighT_Spinodal}
		      \end{center}
	      \end{figure}

	      For T $<$ $2200$ K, we find cavitation in the NVT
              simulations before minima along isotherms are reached,
              due to which we observe a drastic increase in the
              pressure. In an attempt to circumvent this problem, we
              perform restricted ensemble Monte Carlo (REMC)
              simulations. In an REMC simulation, arbitrary bounds
              were imposed on the magnitude of the allowed density
              fluctuations. This is accomplished by dividing the
              simulation box into a number of equal sub-cells and a
              constraint is imposed on the number of particles in each
              of these sub-cells. If the number of sub-cells is $n$,
              the instantaneous number of particles ($N_i$) in each
              sub-cell $i$ is constrained to satisfy ($<N>$-$\delta
              N$) $\le$ $N_i$ $\le$ ($<N>$+$\delta N$), where $<N>$ =
              $(N/n)$ is the average number of particles in each
              sub-cell and $\delta N$ is the maximum allowed deviation
              from the average value. The value of $\delta N$ should
              be greater than unconstrained fluctuation of particles
              which is given by $\Delta N$ = $\left [ <N> k_B T K_T
                \rho \right ]^{\frac{1}{2}}$.  Here $k_B$ is the
              Boltzmann constant, $K_T$ is the compressibility at
              temperature $T$ and density $\rho$. We perform
              REMC simulations in a 1500 particles system by dividing
              the system into 4x4x4 and 5x5x5 sub-cells. The value of
              $\Delta N$ is estimated at different state points based
              on NVT simulations and $\delta N$ is chosen to be three
              times $\Delta N$. The total simulation length in each
              case is 5 million MC cycles. However, we find that even
              in these REMC simulations, the system cavitates, with
              the formation of voids across sub-cell boundaries (with
              each sub-cell satisfying the applied constraint on
              number of particles). Hence we estimate the spinodal at
              these state points from a quadratic extrapolation of
              the isotherms. We fit the
              data points obtained from REMC simulations with a
              quadratic function ($p_0 + a1*(\rho-\rho_0) +
              a2*(\rho-\rho_0)^2$), where $p_0$ and $\rho_0$ are the
              spinodal pressure and density values. The data and the fits are shown in Fig. \ref{fig:LowT_Spinodal}. 

	      \begin{figure}[h]
		      \begin{center}
			      \epsfig{file=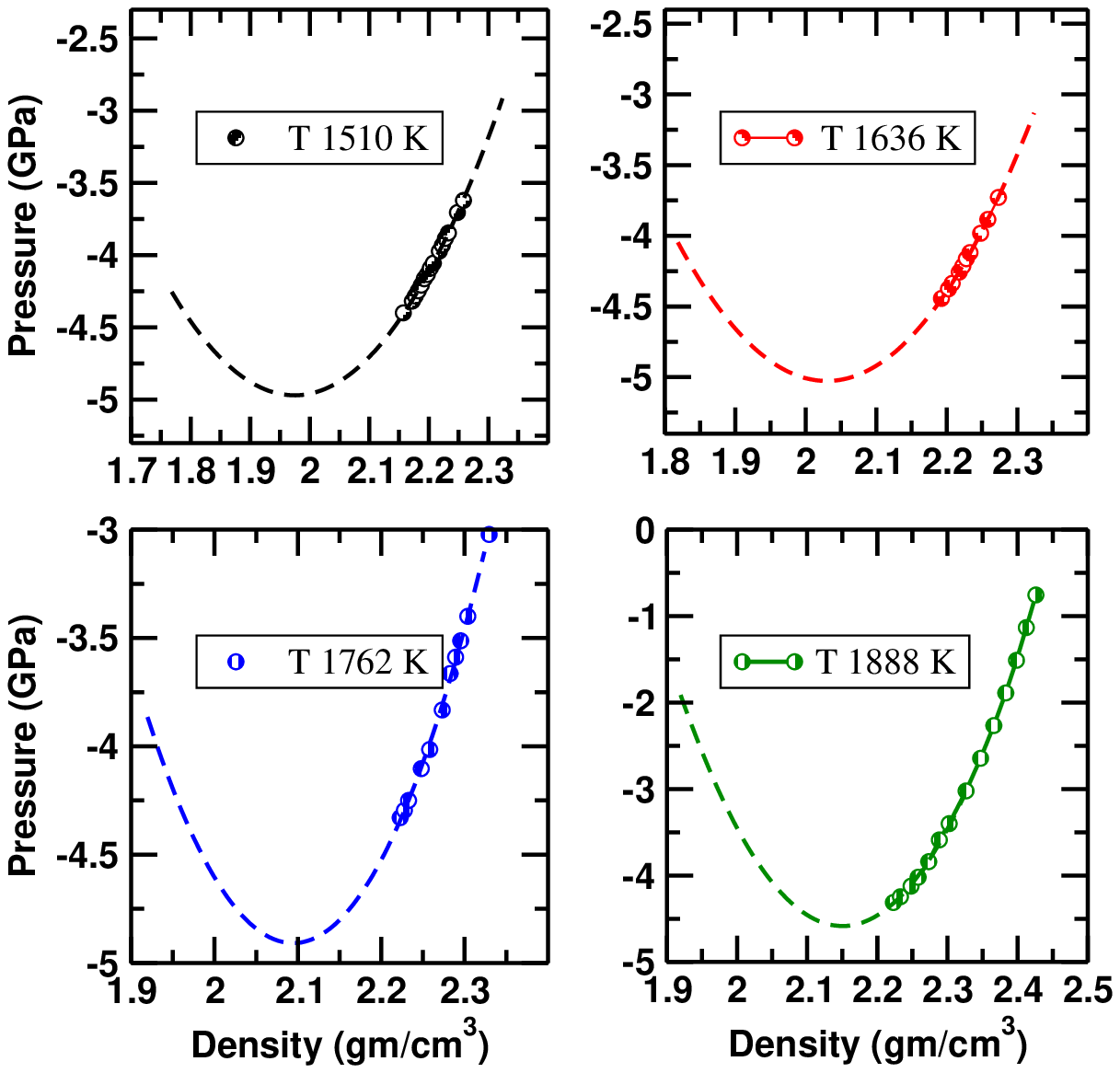,scale=0.7,angle=0,clip=}
			      \caption{The plot of pressure {\it vs. } density for low
			      temperature ($T < 2200 K$) isotherms. The dashed line indicate the quadratic ($p_0 + a1*(\rho-\rho_0) + a2*(\rho-\rho_0)^2$)
			      extrapolation.}
			      \label{fig:LowT_Spinodal}
		      \end{center}
	      \end{figure}

      \item{\bf Tensile limit line: } The locus of maximum tensile stress
        (negative pressure) a system can withstand before it fails 
        defines the tensile limit line. The protocol used to
        obtain the tensile limit is as follows. At a given temperature
        we first equilibrate the system at a high pressure (for T $<$
        $1510$ K we equilibrate at P $=$ $-2.26$ GPa and for T $>$
        $1510$ K at P $=$ $0$ GPa, by performing NPT MD simulation for
        $4$ ns. All the simulations are performed with 512
        particles). We then apply a tensile pressure to
        this equilibrated sample that increases at a specified rate. We perform simulations at
        two different constant rates of change of tensile
        pressure. When the system reaches its limit of tensile
        strength, the system's density decreases drastically towards zero. In
        Fig. \ref{fig:Slow_Rate}, and Fig. \ref{fig:Fast_Rate} we
        show the applied pressure against the measured density for a range of temperatures, from
        which we obtain the tensile limit line.  At faster stretching
        rate (10 MPa/ps) we find tensile limits that are consistent
        with the spinodal estimates described above. For slow stretching rate (0.1
        MPa/ps) we find that the system cavitates at higher pressure
        values. We have also performed simulations at intermediate
        rate, 1.0 MPa/ps (not reported in this letter) and we find
        that the estimated tensile limit line lies in between the estimates
        obtained from the fast and slow stretching rates. For slow
        stretching rate we perform simulations for around $70$
        $ns$. These simulations are performed with the LAAMPS package. 

	      \begin{figure}[h]
		      \begin{center}
			      \epsfig{file=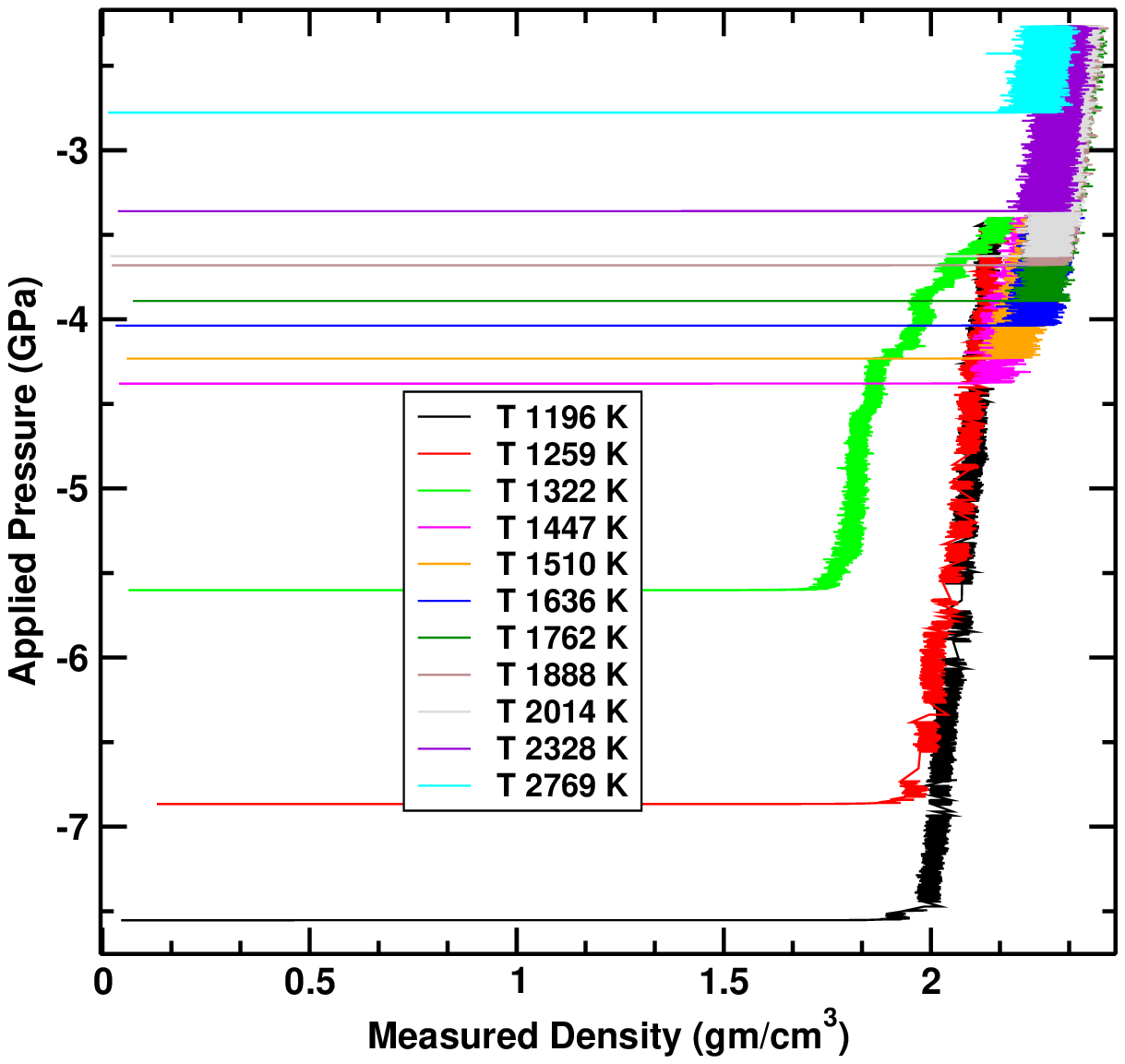,scale=0.5,angle=0,clip=}
			      \caption{The plot of applied pressure against measured density for different
			      isotherms. The stretching rate is 0.1 MPa/ps}
			      \label{fig:Slow_Rate}
		      \end{center}
	      \end{figure}

\clearpage
\newpage
	      \begin{figure}[h]
		      \begin{center}
			      \epsfig{file=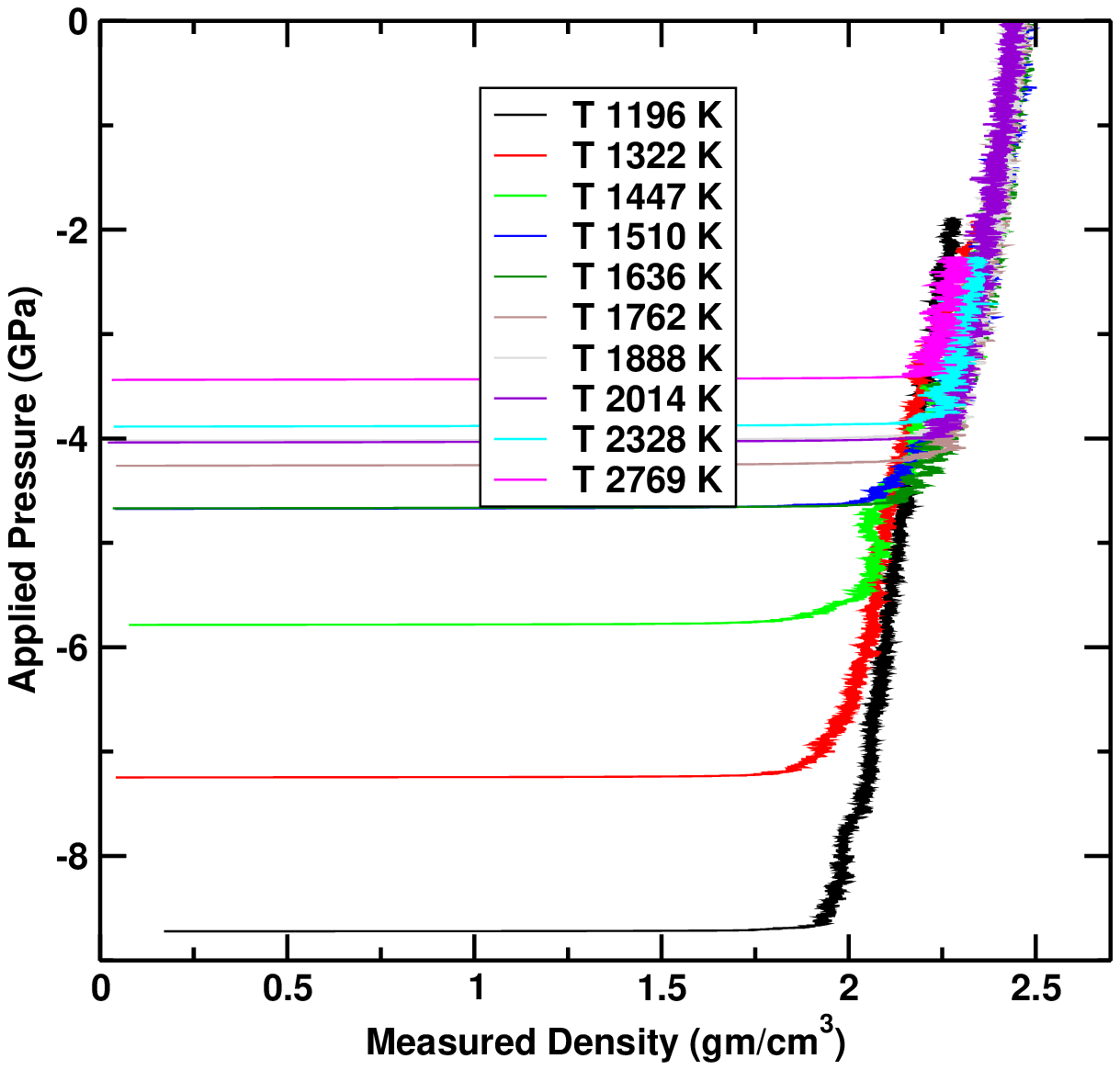,scale=0.5,angle=0,clip=}
			      \caption{The plot of applied pressure against measured density for different
			      isotherms. The stretching rate is 10.0 MPa/ps}
			      \label{fig:Fast_Rate}
		      \end{center}
	      \end{figure}

	\item {\bf Temperature of maximum density:} The temperature of
          maximum density is the locus of isobaric maxima of density
          $\rho$ {\it vs.} $T$ ($(\partial \rho/\partial T)_P = 0$) or
          the locus of isochoric minima of pressure $P$ {\it vs.} $T$
          ($(\partial P/\partial T)_V = 0$). For pressure values above
          $P$ = $-3.80$ GPa, we obtain the TMD from NPT MD simulations
          to locate isobaric maxima of density. But below $P$ =
          $-3.80$, since cavitation is a possibility in NPT
          simulations, we perform NVT MD simulations to locate
          isochoric minima of pressure. The TMD obtained from density
          maxima along isobars and pressure minima along isochores are
          shown in Fig. \ref{fig:TMD_Isobar} and
          Fig. \ref{fig:TMD_Isochore} respectively. The simulation
          length for these state points is around $10$ ns (more than
          $100 \tau$ where $\tau$ is the relaxation time) and the
          system size is 512 particles.

	      \begin{figure}[h]
		      \begin{center}
			      \epsfig{file=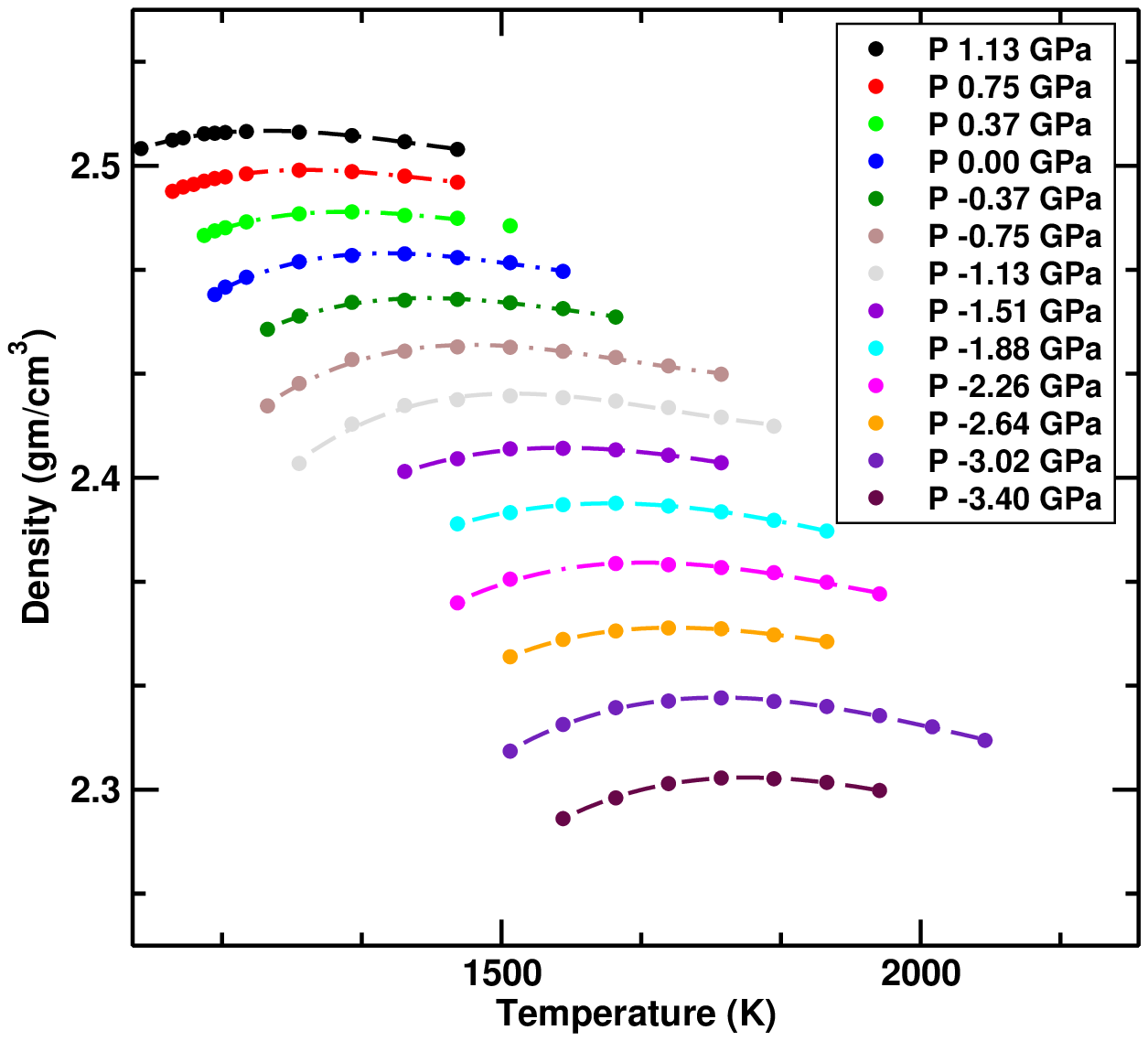,scale=0.43,angle=0,clip=}
			      \caption{The plot of density against temperature for different isobars. The
			      maxima obtained at each isobar constitutes the TMD line.}
			      \label{fig:TMD_Isobar}
		      \end{center}
	      \end{figure}

	      \begin{figure}[h]
		      \begin{center}
			      \epsfig{file=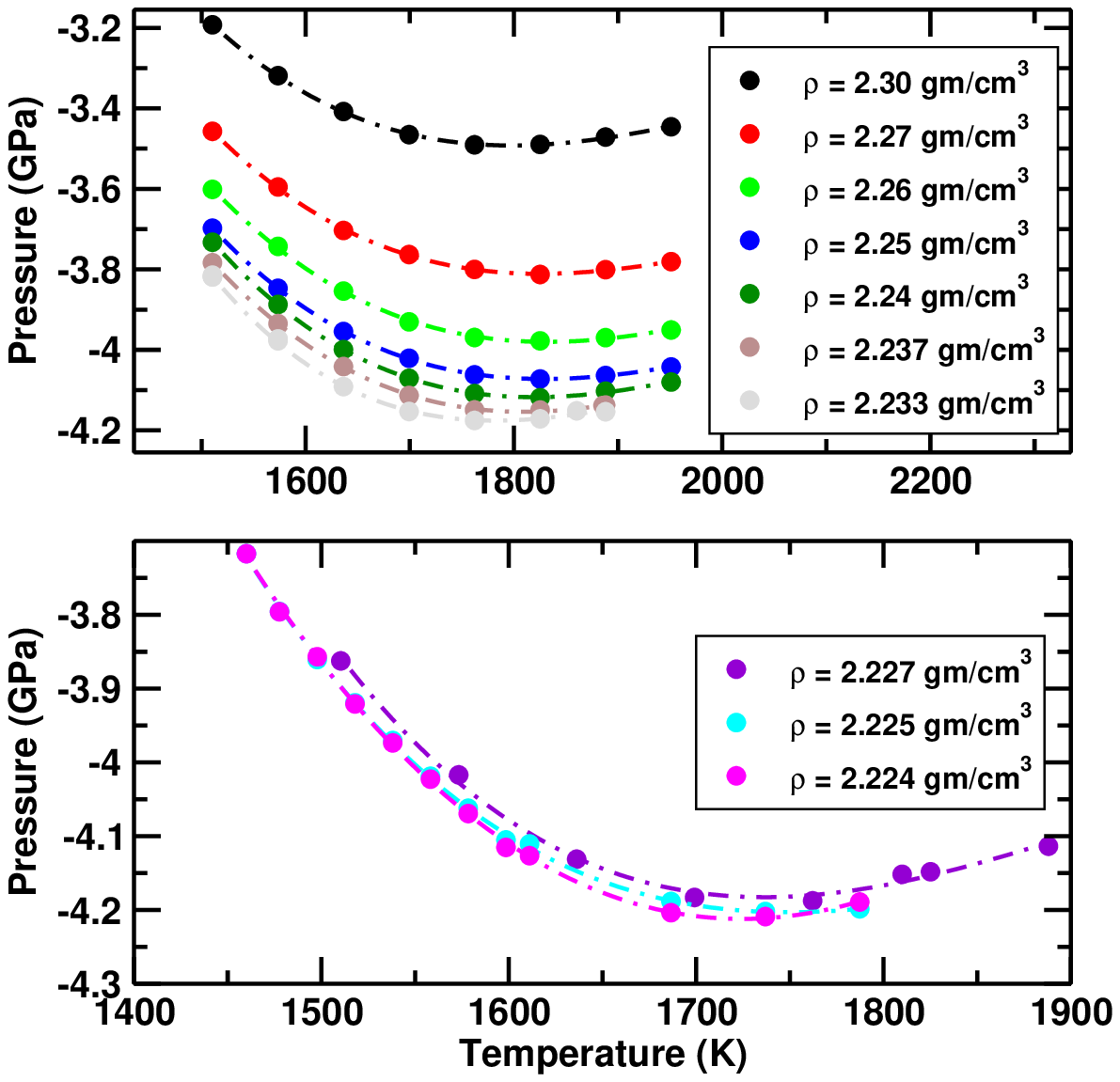,scale=0.43,angle=0,clip=}
			      \caption{(top panel) The plot of pressure against temperature for different isochores. The
			      minima obtained at each isochore constitutes the TMD line. In the bottom panel we show last 
			      three isochores at which we could perform NVT simulations. Below these densities, the system
			      cavitates before going through a pressure minimum.}
			      \label{fig:TMD_Isochore}
		      \end{center}
	      \end{figure}

	\item {\bf Temperature of minimum density:} TMinD is locus of
          density minima points, crossing which, the system returns to
          normal behaviour. Finding the TMinD line in supercooled
          silicon is challenging since we must simulate the system
          deep inside the supercooled region of the phase diagram
          (where crystallisation, slow equilibration and cavitation
          pos hurdles to obtaining equilibrated data). In order to
          obtain equilibrated data we employ NPT parallel tempering
          Monte Carlo simulations ({\it see Understanding Molecular
            Simulation by D. Frenkel and B. Smit}), in which we
          perform swaps between replicas at different T and P. The
          paths chosen to perform parallel tempering are shown in the
          Fig. \ref{fig:TMinD_PT}.  We perform simulations,
          with a system size of 512 particles, up to 60 million MC
          cycles. To check the equilibration of the system using the
          following tests: We ensure that each replica gets swapped up
          and down in temperature T alteast 100 times. We calculate the
          mean squared displacement (MSD) for each replica and
          ascertain that every particle in the system has moved
          atleast $5$ $\sigma$. 
	      
	      \begin{figure}[h]
		      \begin{center}
			      \epsfig{file=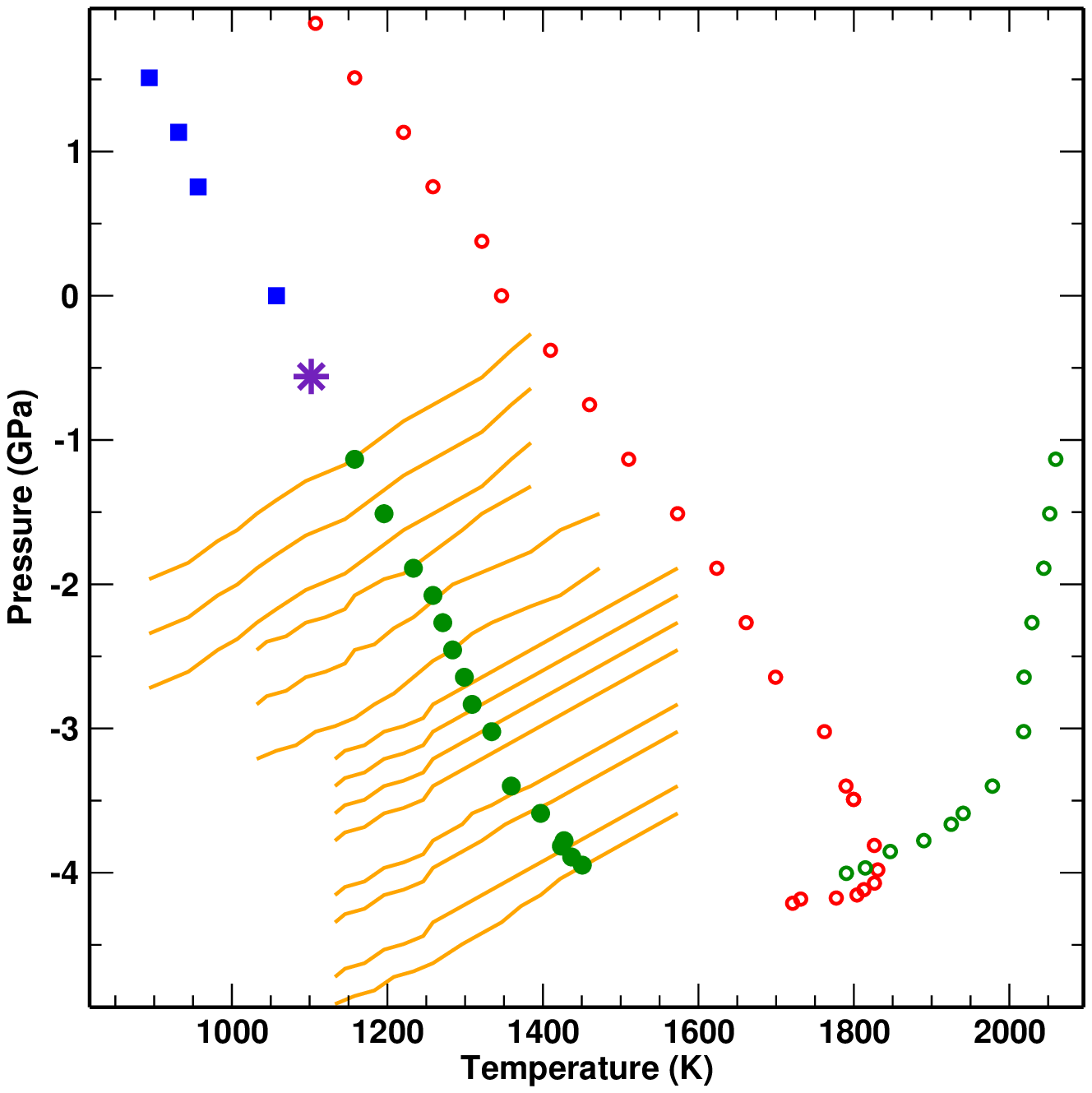,scale=0.5,angle=0,clip=}
			      \caption{The phase diagram showing the paths chosen to perform NPT-PT simulations.
			      The orange lines are the paths along which NPT-PT simulations are performed. The green
                              points indicate compressibility extrema, the red points the density maxima, the purple points the liquid-liquid coexistence line, and the star indicates the location of the critical point.}
			      \label{fig:TMinD_PT}
		      \end{center}
	      \end{figure}

	      \begin{figure}[h]
		      \begin{center}
			      \epsfig{file=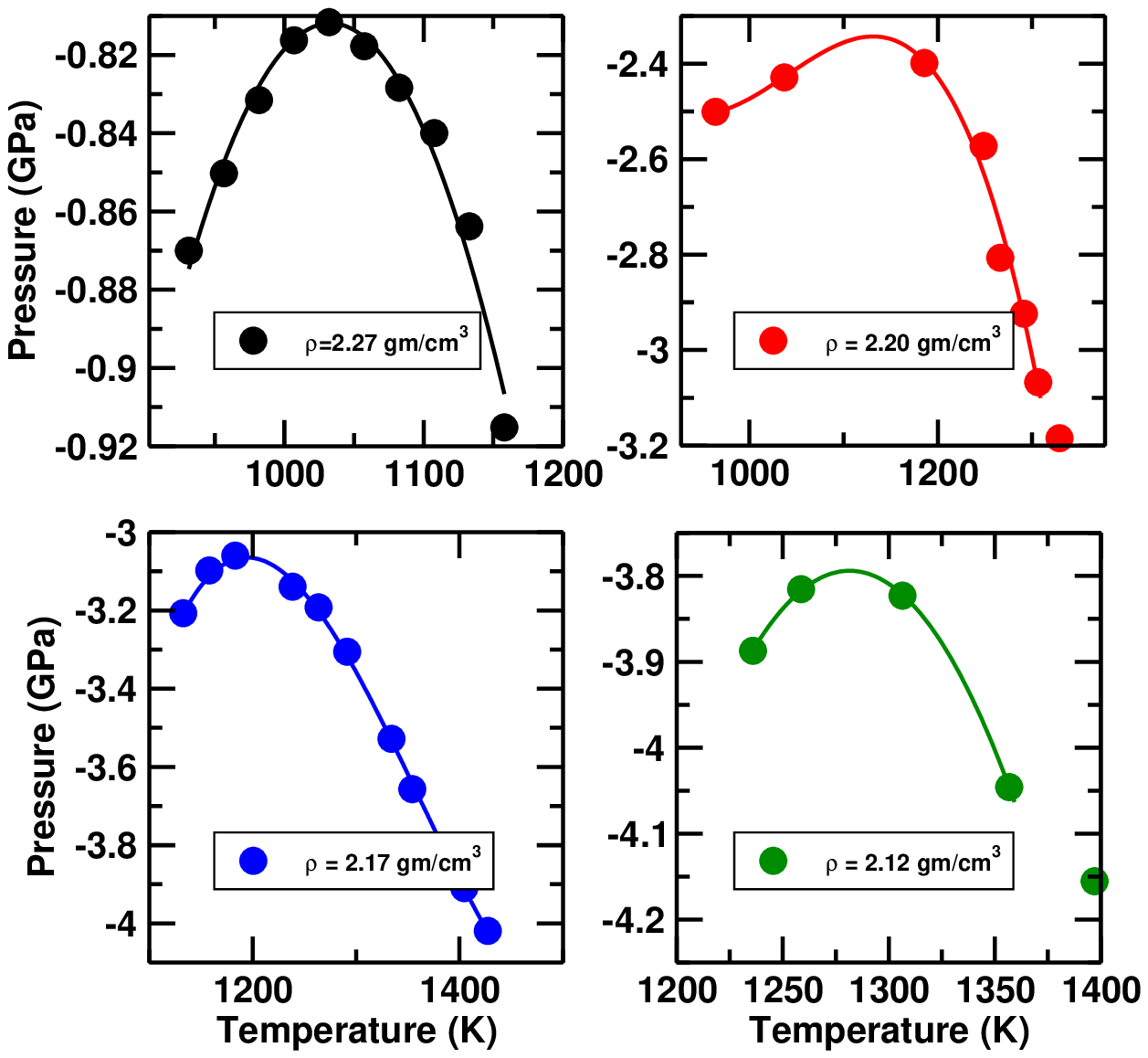,scale=0.5,angle=0,clip=}
			      \caption{Pressure {\it vs.} temperature for different isochores. The
			      maxima obtained at each isochore constitute the TMinD line.}
			      \label{fig:TMinD_Isochore}
		      \end{center}
	      \end{figure}

\clearpage
\newpage

        \item {\bf Temperature of minimum compressibility (TMinC):} We
          perform NPT MD simulations (with 512 particles) to obtain
          TMinC points. The relaxation time at these state points are
          of the order of a few ns (or 10 to 30 million MD steps). We
          face the issue of cavitation below $-3.80$ GPa.  In this
          region of the phase diagram we have performed simulations for
          a minimum of 10 independent samples to construct the equation of state (EOS), from
          which we obtain the compressibility by taking the derivative
          of the pressure with respect to density.

	      \begin{figure}[h]
		      \begin{center}
			      \epsfig{file=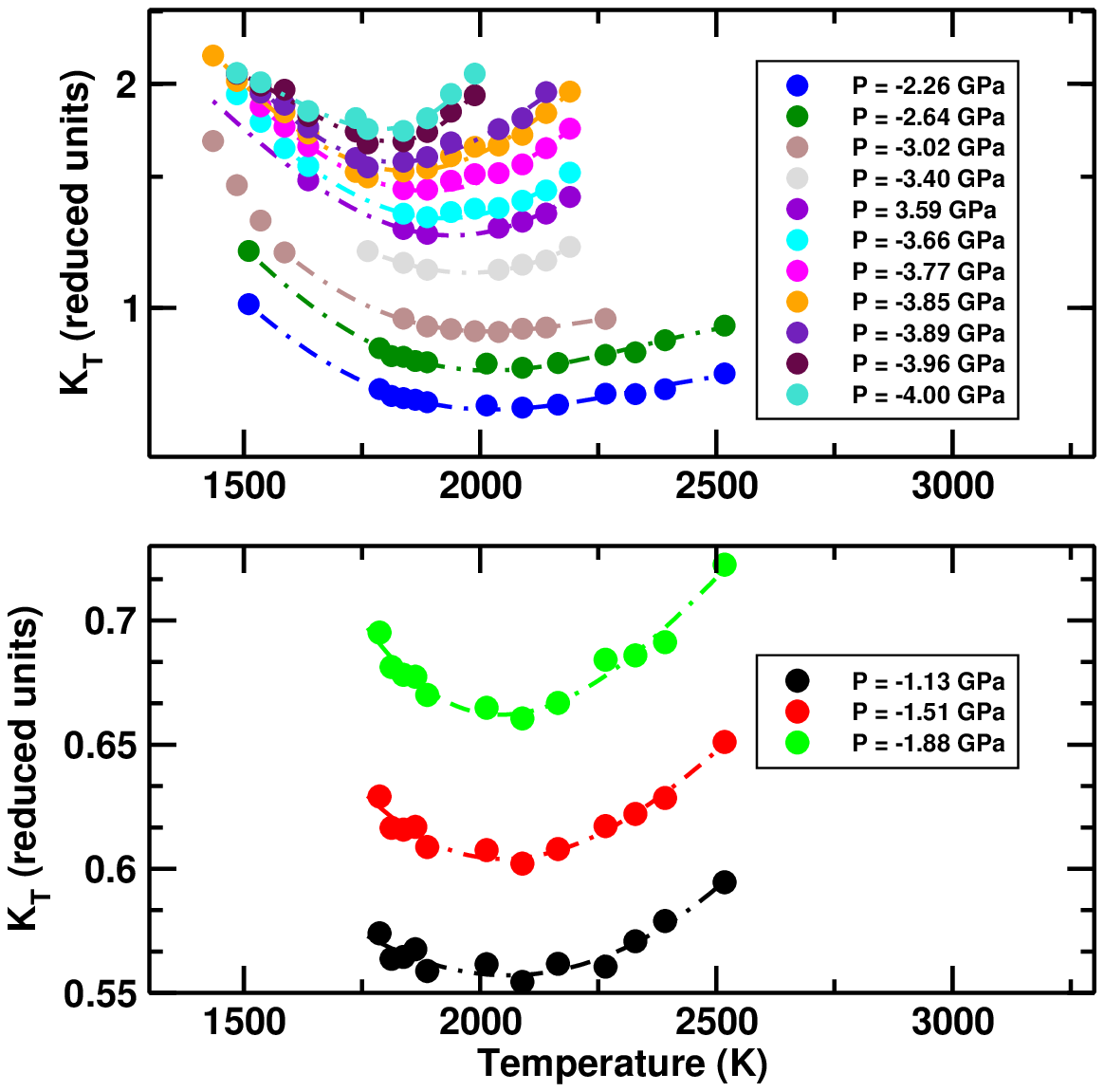,scale=0.5,angle=0,clip=}
			      \caption{Compressibility {\it vs.} temperature for different isobars. The
			      minima obtained along isobars constitute the TMinC line.}
			      \label{fig:TMinC_Isobar}
		      \end{center}
	      \end{figure}

        \item {\bf Temperature of maximum compressibility:} The high
          pressure (P $>$ $-2$ GPa) compressibility maxima
          ($K_T^{max}$) are shown in the manuscript (Fig. 2).
          Compressibility data from which $K_T^{max}$ are obtained for
          P $<$ $-2$ GPa are shown in the
          Fig. \ref{fig:TMaxC_Isobar}. We find that as the system
          crosses the $K_T^{max}$ line from high T to low T (at a
          chosen pressure value), the relaxation times increases from
          picoseconds to tens of nanoseconds. Nearing the
          compressibility maxima we also find that many samples
          crystallise. The $K_T$ values shown in the
          Fig. \ref{fig:TMaxC_Isobar} are calculated from both volume
          fluctuations measured in NPT MD simulations and from
          derivatives of pressure from NVT MD simulations. For
          pressure value below $-3.90$ GPa, the system cavitates easily and hence 
          we cannot evaluate the location of $K_T^{max}$ in this
          region of the phase diagram.

	      \begin{figure}[h]
		      \begin{center}
			      \epsfig{file=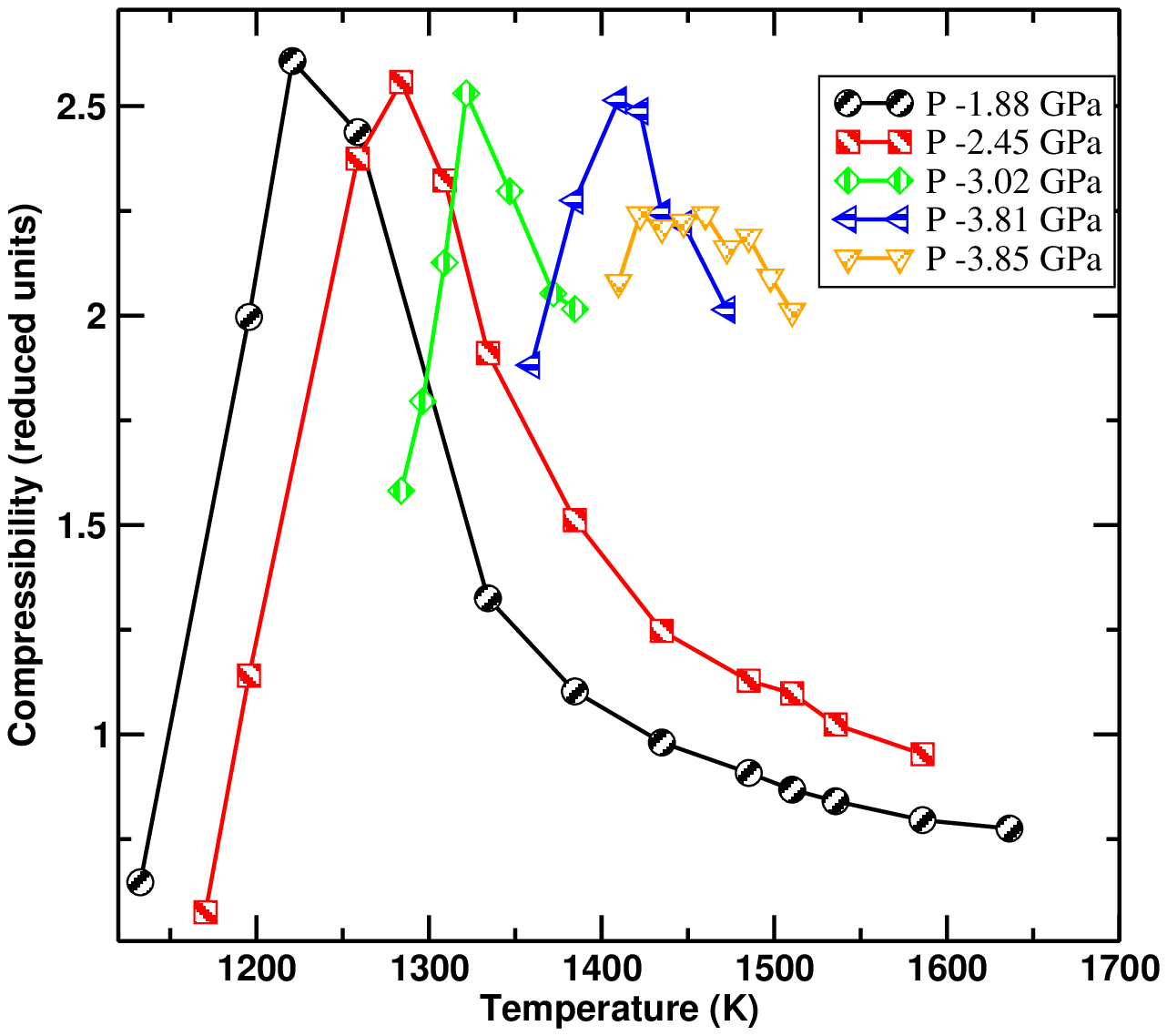,scale=0.5,angle=0,clip=}
			      \caption{Compressibility {\it vs.} temperature for different isobars. The
			      maxima obtained along isobars constitute the TMC line.}
			      \label{fig:TMaxC_Isobar}
		      \end{center}
	      \end{figure}

\end{enumerate}